\documentclass[options]{acmart}

\makeatletter
\def\@ACM@checkaffil{
    \if@ACM@instpresent\else
    \ClassWarningNoLine{\@classname}{No institution present for an affiliation}%
    \fi
    \if@ACM@citypresent\else
    \ClassWarningNoLine{\@classname}{No city present for an affiliation}%
    \fi
    \if@ACM@countrypresent\else
        \ClassWarningNoLine{\@classname}{No country present for an affiliation}%
    \fi
}

\setcopyright{acmcopyright}
\copyrightyear{2023}
\acmYear{2023}
\acmDOI{XXXXXXX.XXXXXXX}
\acmPrice{15.00}
\acmISBN{978-1-4503-XXXX-X/18/06}

\usepackage{subfigure}
\usepackage{graphicx}
\usepackage{listings}
\usepackage{algorithm}
\usepackage{algorithmic}
\usepackage{xcolor}         
\usepackage{amsmath}
\usepackage{hyperref}       
\usepackage{url}            
\usepackage{booktabs}       
\usepackage{amsfonts}       
\usepackage{nicefrac}       
\usepackage{microtype}      
\usepackage{lipsum}

\begin{document}

\title{Learning Personalized User Preference from Cold Start in Multi-turn Conversations}

\author{Deguang Kong, Abhay Jha and Lei Yun}

\email{doogkong@gmail.com, jhaabha@amazon.com, yleimzn@amazon.com}
\authornote{The work was done while authors were working at Amazon Alexa AI. Correspondence to: Deguang Kong<doogkong@gmail.com>.}

\affiliation{Amazon Alexa AI, Sunnyvale, CA, U.S.A}

\renewcommand{\shortauthors}{D.Kong and  et al.}

\begin{abstract}
This paper presents a novel teachable conversation interaction system (TAI for short) that is capable of learning users’ preferences from cold start by gradually adapting to  personal preferences. In particular, the TAI system is able to automatically identify and label users’ preference in
live interactions, manage dialogue flows for interactive teaching sessions, and reuse learned preference for preference elicitation. We develop the TAI system by leveraging BERT encoder models to encode both dialogue and relevant context information, and build action prediction (AP), argument filling (AF) and named entity recognition (NER) models to understand the teaching session. We adopt a
seeker-provider interaction loop mechanism to generate diverse dialogues from cold-start. TAI is capable of learning users' preference, which achieves 91.22\% turn level accuracy on out-of-sample dataset, and has been successfully adopted in production.
\end{abstract}




\maketitle

\section{Introduction}

In conversation AI~\cite{NEURIPS2020_e9462095}, the interaction between users and conversation agents is helpful to guide users to achieve certain tasks. Customers may desire and expect the conversation agent to provide proactive personalized experiences in order to fulfill an advertised promise of a helpful, personal assistant~\cite{labutov-etal-2018-lia}. One significant effort would be accelerating personalization  by harmonizing disparate preference learning experiences across the conversation agent.


 
In this paper, we present a teachable conversation agent system that 
augments the task oriented AI agents with an interactive teaching capability. 
In particular, the \emph{challenges} we are facing here include \emph{(C1)} how to allow users to naturally  start  with  describing  the preferences  and  its  conditionals at a high-level, and then recursively teach the agent through the conversations, \emph{(C2)} how to understand diverse related entities and conversation flow users may have for their preferences in a multi-turn dialog, \emph{(C3)} how to train the conversation model given scare training data in cold-start environment, \emph{(C4)} how to reuse the previously taught user preferences across different domains,  {\it etc}.


\begin{figure}
\centering
  \includegraphics[
  width=9cm,height=3.5cm]{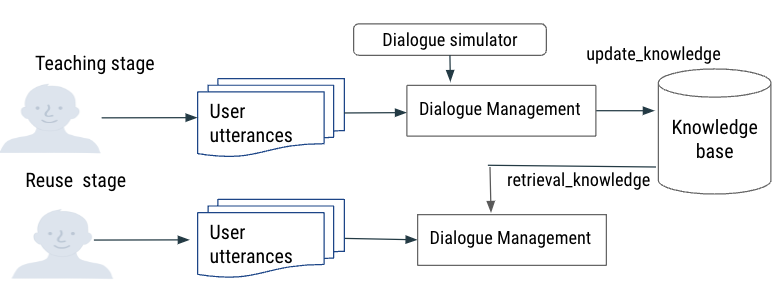}
  \caption{The working flow of learning user preferences conversational system (including teaching stage and reuse stage).}
  \label{fig:teaching_reuse}
\end{figure}
\begin{figure*}[h!]
\centering
  \includegraphics[scale=0.3]{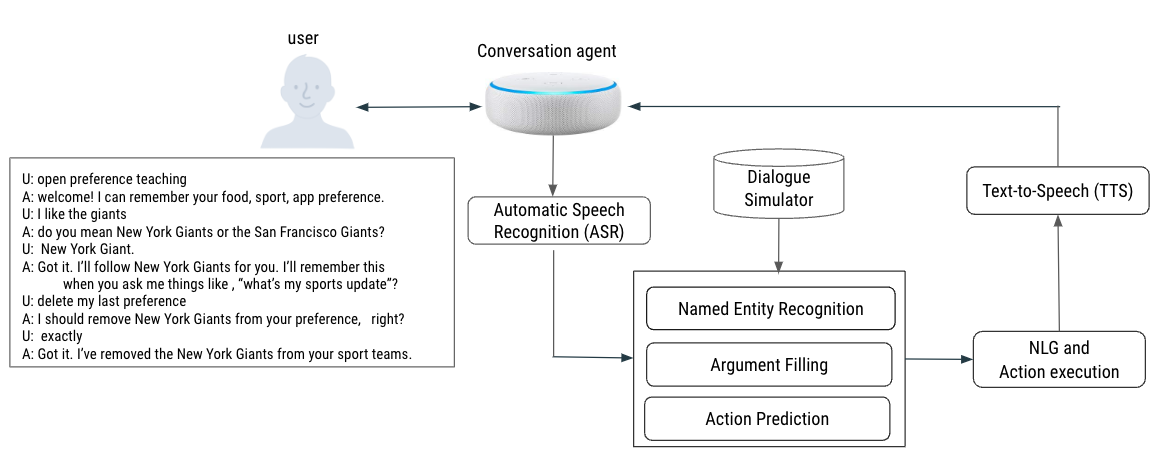}
  \caption{Overview of conversation understanding and dialogue management}
  \label{fig:conversation_module}
\end{figure*}
Despite some works~\cite{10.5555/1619797.1619888,DBLP:conf/uist/LiRJSMM19,DBLP:conf/acl/YinN17,icassp/7953253,DBLP:conf/iwsds/HouWYYHS18,DBLP:conf/interspeech/KobayashiYIF19,DBLP:conf/acl/XuH18,DBLP:conf/acl/zhao-feng-2018-improving,DBLP:journals/corr/abs-2005-03923,DBLP:journals/corr/abs-1902-10909, DBLP:conf/naacl/GooGHHCHC18, DBLP:conf/EMNLP/takanobu2019guided, DBLP:journals/corr/abs-1712-02838} 
about learning from user interactions, they are not sufficient to address the above challenges. For example, interaction learning~\cite{DBLP:conf/acl/WangGLM17} is widely used in learning visual scene concept~\cite{Mao2019NeuroSymbolic} about attributes (such as color, shape, etc), learning semantic concept (e.g., "set my room to cozy")~\cite{Ping2020} using concept parser, as well as  games~\cite{ijcai2019-844} using the symbolic execution for state progression and scene understanding. However, the method is not directly applicable for preference teaching. While reinforcement learning approach~\cite{DBLP:journals/corr/abs-1709-06136} views the dialogue conversation as a sequential decision
making process that learns optimal policy, it is over-complicated and difficult to customize for different domain teaching sessions.


To our knowledge, this work is the \emph{first} work that builds a conversation agent which allows the users to share their preferences across different topics and get more relevant and personalized results. The \emph{contribution} of this paper is summarized as follows. 

    
{\bf $\bullet$} To address \emph{challenge C1}, a multi-turn domain agnostic dialog system is designed to support user initialized teaching, allowing users to describe the preference,  clarify ambiguities from explicit  instructions.

{\bf $\bullet$} To address \emph{challenge C2},  the conversation agent  performs a multi-turn action prediction loop upon named entity recognition, action prediction  and argument filling models by encoding dialog context using BERT and Seq2Seq model pipeline.  
    
{\bf $\bullet$} To address \emph{challenge C3}, we designed a dialogue simulator module to generate synthetic training data with adequate annotations guided by the seeker-provider interaction loop to achieve the seeker goals from cold-start. 

{\bf $\bullet$} To address \emph{challenge C4}, we enable centralized preference management, which stores the taught preference  in  a  persistent  knowledge  base,  facilitating  effective  reusability  and  generalizability  of learned knowledge. 

    
\section{System Overview}

\label{Section:system_overview}



TAI  consists of two stages (i.e., teaching stage and reuse stage).  In the teaching stage, given user utterance, TAI will call a dialogue management module to learn users’ preferences from interaction. Users’ preferences will be stored at the knowledge base after standardizing the schema for future use.  After enabling reuse,  when a user asks the specific questions, the domain will retrieve the relevant preference (stored in the knowledge base) and further infer additional latent customer preferences to curate more proactive personalized experiences to users. Fig.~\ref{fig:teaching_reuse} depicts how the TAI system works. The purpose of dialogue management (Fig.~\ref{fig:conversation_module}) is to conduct interactive sessions with users for learning preferences and drive the conversations based on the seeker goals to complete the objectives. Essentially, it consists of several components:

{\bf $\bullet$ Preference parser}:  Conversational agent will bridge the prediction gap to understand and interpret the preference for the given utterance from the user. For example, it will understand the slot relevant to preference in utterance such as \texttt{I love Giants [sportsteam]}. This is mainly achieved by NER illustrated in  Section~\ref{Section:multi_turn_action_prediction_loop}. 

{\bf $\bullet$ Dialogue management}: After the session has been initiated by the user in the conversation, the dialog management module will predict the right clarification question to ask the user, based on the context of utterance and the interactions. Then the user will respond for clarification (if needed) and continue the multi-turn conversations until the respective preferences are grounded before storage of them for future re-use. This is achieved by a multi-turn action prediction loop illustrated in Section~\ref{Section:model_breakdown}.

{\bf $\bullet$ Dialogue simulation from cold start}: we build the simulator based on seeker provider interaction loop to generate diverse dialogues that are beyond these seed examples by enriching them with surface form (i.e. utterance) variations, other dialogue flows and possible unhappy cases. This is mainly achieved by simulation in Section~\ref{Section:simulator_data_generation}.

{\bf $\bullet$ Centralized Knowledge Storage}: The taught preference is stored in a persistent knowledge graph, which facilitates effective reusability and generalizability of learned knowledge. In particular, it supports the following functionalities:

\begin{lstlisting}[frame=single]  % Start your code-block
Update_KB(User_Id, &PrefList) \\
Retrieve_KB(User_Id, &PrefList)
\end{lstlisting}
\begin{lstlisting}[frame=single]  % Start your code-block
Retrieve_KB(User_Id, &PrefList)
Update_KB(User_Id, &PrefList) 
\end{lstlisting}
which enables the TAI to store, update and retrieve the \texttt{[user\_id, preference\_list]} pairs in the knowledge base. 
Through these interactions, it leads to more complex personal interactions and action requests that surface more nuanced gaps in conversation understanding.

\subsection{Utterance Examples}
\label{sec:appendix_golden}

The sampled utterance example is shown in Table~\ref{tab:goldens}.

  \begin{table*}[ht]
  \small 
 \caption{Sample golden utterances}
  \centering
  \begin{tabular}{lcc }
    \toprule
   Domain   &  Goal & Utterances    \\
    \midrule
  sports        & like sport team &  The Yankees are my favorite baseball team   \\
    sports        & like sport team & My favorite teams are the Yankees and The Knicks  \\
      sports        & like sport team & Add the Warriors to my favorites  \\
        sports        & like sport team &  I am a new York giant fan.   \\
    sports        & dislike sport team & Remove the Yankees from my favorite sports teams  \\
    sports        & dislike sport team & Unfollow San Francisco giant.   \\
    restaurant    & restaurant/cuisine one loves & I like to dine at Thai restaurant \\
    restaurant    & restaurant/cuisine one loves & I enjoy seafood restaurant. \\ 
    preference management & reset preference & Forget everything you've learned about my preferences \\
    preference management & reset preference  & Delete all my preferences \\
    preference management & delete preference & Remove Mexican food from our food preferences. \\ 
    preference management & delete preference  & I want you to forget my preference for vegetarian food. \\
    weather provider & conditional preference  & I prefer big sky for weather update. \\
    weather provider & conditional preference  & Always use RADAR for weather forecasting. \\
    \bottomrule
  \end{tabular}
  \label{tab:goldens}
\end{table*}

\section{Simulating Dialogues to Address Cold Start Issues}
\label{Section:simulator_data_generation}

We design and implement the dialogue simulator module (``simulator" for short) to generate synthetic training dialogues with adequate annotations since we do not have any user data before product launch. 
For this purpose, the \textbf{input} to the simulator module would be small number of annotated seed dialogues with descriptions, and the \textbf{output} would be diverse dialogues that are generated beyond these seed examples by enriching them with surface form (i.e. utterance) variations, other dialogue flows and possible unhappy cases. In the conversation, the user is annotated as a \texttt{seeker} who seeks to accomplish a goal. The conversation agent (annotated as a \texttt{provider}) provides information or a service to help the seeker complete the goal. The simulation is mainly achieved by {\bf seek-provider interaction}. 

\begin{figure}[h!]
\centering
  \includegraphics[scale=0.35]{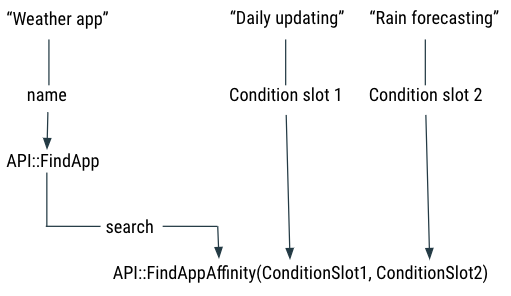}
  \caption{An example of entity transfer graph for the utterance \texttt{I prefer the weather app for daily updating and rain forecasting}}
  \label{fig:entity_graph}
  \vspace{-0.5em}
\end{figure}

{\bf Stage 1: Seeker goal} In this stage, a fixed seeker goal is sampled from the goal sampler. To describe the seeker goal of a conversation that consists of a sequence of API calls by the provider to fulfill the seeker's goal,  we adopt  an \emph{entity transfer graph}, i.e., a \emph{directed acyclic graph (DAG)} where the vertices consists of the entities supplied by the seeker and the API calls of the provider, and the edges encode the way that seeker entities and entities returned by API calls are transferred (i.e. carried over), such that the arguments of each API call in the goal are therefore encoded as parents in the graph. Fig.~\ref{fig:entity_graph} gives an example of entity transfer graph.  It draws an API by following a Markov chain distribution (denoted as $A_{1:i}$ for API sequences [1:i]) estimated from the seed conversation, i.e, 
\begin{equation}
   \Pr(A_{1:i}) = \prod_j 
   \Pr(A_i| A_{j|{j<i}})
\end{equation}
where the probability of goal $\Pr(A_{1:j})$ is achieved by sampling a sequence of intents (via APIs) in the sequential order (with index $i$, $j$).

Then starting with an empty entity transfer graph, seeker goals are sampled incrementally by repeatedly determining if a new vertex should be added. In particular, it draws an API by following a Markov chain distribution estimated from the seed conversation, i.e, 
\begin{equation}
    Pr(\text{goal}) = \Pi_i Pr(\text{API}_i| 
\end{equation}
where the probability of goal $Pr(\text{goal})$ is achieved by sampling a sequence of intents (via APIs) in the sequential order (e.g., $i$, $j$). The sampling sequence $j$ is restricted to be $i-1$ for Markov sampling. For each API argument, it randomly transfers existing entities or elicits new entities from the seeker, and then adds the entities elicited (if any) and adds the new API vertex with parent edges from the transferred entities.

{\bf Stage 2: Seeker-Provider Interaction Loop} The provider and seeker agents exchange a series of messages composed of their respective actions and utterances. On one hand, the seeker incrementally reveals the goal to the provider with actions corresponding to dialogue acts. On the other hand, the provider collects all the arguments necessary to make API calls, makes those calls and informs the seeker of the results with NLG template actions and API calls. Starting with the seeker, the conversation agent executes at each turn for a) action sampling (i.e., a sequence of actions is drawn according to a distribution of possible actions), b) NLG: an utterance corresponding to the dialogue acts associated to the sampled actions is drawn. Both parts provide the necessary annotations.

\subsection{Design principle}
To dive deep into the simulator mechanics, the design is guided by \textbf{generating  diverse dialogues}. It could be generated in the follow ways. 

First, dialog variations can be introduced by resampling the entities of the conversation using catalogs and using alternative utterance wordings. For example, the slot \texttt{sport\_team} in ``I love sport\_team'' would be substituted by other entities in catalog.  Second, the user could change the information provided in prompts in each turn. For example, when a user says "I don't like the Gaint", the generated clarification would be "which gaint do you mean, New York Gaint or San Francisco gaint?" The benefit is to proactive suggest users' preferences for feature discovery. Third,  it is important to simulate cases with sufficient randomness or even errors, since neither the conversation agent or the user will behave optimally. For example, when a user says "what's my sports update?", the conversation agents may see the situation that a user has not set  his preference yet, and asked "Please set up your sport preference before getting notifications."  The clarification sub-dialogues is to reduce uncertainty in conversation agents' estimation of the dialogue state.  Finally, to avoid overfitting during model training, it is important to generate a previously unseen conversation to ``realistic" simulator potential errors or misunderstandings. 

\subsection{Architecture}

Based on the above design principle, we illustrate the components of the dialogue simulator and how they interact with the modeling component.

The simulator adopts the concept of dialogue acts to express the semantics associated to utterances. For example, \texttt{inform} is the act used by the seeker to communicate their intent and by the provider to describe the result of their actions. Dialogue acts are used at the interface between the natural language generation (NLG) unit, which converts the dialogue content and direction to a corresponding language realization, and conversation agent.

\textbf{Procedure} Natural Language Generation component is currently based on templates and is responsible for the generation of the seeker's utterances from dialogue acts, where seeker utterances are first categorized with Dialogue Act classifier and then converted into NLG templates using simple heuristics. In particular, it consists of three key steps: 

{\bf Step 1:} A small batch of ~100 dialogues is simulated with the extracted seeker NLG templates.  

{\bf Step 2:} A Mechanical Turk task is submitted to obtain variations of the extracted seeker NLG templates by crowdsourcing, where paraphrasing is done both at the utterance level and conversation level. 

{\bf Step 3:} A larger set (e.g., 50k) of annotated dialogues is simulated and used for supervised training the conversation model.

\subsection{Simulated dialogues for addressing cold-start issues}
\label{section:simulator}

An example of simulated dialogue for addressed cold-start issue is shown below.

\begin{algorithm}
\small
\begin{flushleft}
\label{alg:alg1}
{\bf U-1}: "type": "User\_intent", "[String] User side utterance": "forget everything I have taught you", "[List of string] UserNLGs associated with the user utterance": preferenceteaching.events.DeleteAllAffinityEvent();

{\bf U-2}:"type": "User\_intent", "partially\_normalized\_value": "[let it be|SINGLESITE.Confirmation -> confirmation1]", "fully\_normalized\_value": ["com.cu.preferenceteaching.events.ConfirmationEvent (confirmation=confirmation1)"]; 

{\bf S-1}: "type": "api", "[String] User side utterance": "getAllAffinityAction() -> getAllPreferenceResult1",", "[List of string] UserNLGs associated with the user utterance": getAllAffinityAction() -> getAllPreferenceResult1"; 

{\bf S-2}:  "type": "nlg", "notify\_com.cu.preferenceteaching.actions.getAllAffinityAction\_success (getAllAffinityResult=getAllPreferenceResult1)";

{\bf S-3}:  "type": "sys", "normalized\_value": "wait\_for\_user\_input()";

{\bf S-4}: "type": "api", "normalized\_value": "com.cu.preferenceteaching.actions.deleteAllAffinityAction  (confirmAction=confirmation1) -> deleteAllPreferenceResult1";

{\bf S-5}: "type": "nlg", "fully\_normalized\_value": "notify\_com.cu.preferenceteaching.actions.deleteAllAffinityAction \_success (deleteAllAffinityResult=deleteAllPreferenceResult1). 



\end{flushleft}
\end{algorithm}

\subsection{Preference teaching use case} 

In the context of preference teaching, the user goal in the goal-oriented dialogue system is a sequence of APIs (such as \\
\texttt{setDietOrCuisineAffinity}, 
\texttt{setSportAffinity}) and their corresponding arguments and values. One can sample the dialogue based on predefined templates and perform sampling on the entities (e.g., \texttt{I love \{entity\}}). The arguments and their values of APIs are sampled as well for more variations. We perform sampling of API transitions in dialogue flows. For example, when computing API transition weight, we consider factors such as API transitions observed in dialog examples provided by the developer, shared entities amongst APIs, input-output relation between APIs, {\it etc}. We then combine these weights using the mixing ratio, treated as hyper parameters tuned for simulated dialogue generation. An example of simulated dialog annotated with markup language (where U- gives the user side utterance and S- gives the conversation agent side actions) is illustrated in ~\ref{section:simulator}. 




\section{Mutli-turn Action Prediction Loop}
\label{Section:prediction_modeling}

\subsection{Overview}
\label{Section:multi_turn_action_prediction_loop}
\begin{figure}[t]
\centering
  \includegraphics[
  scale=0.32]{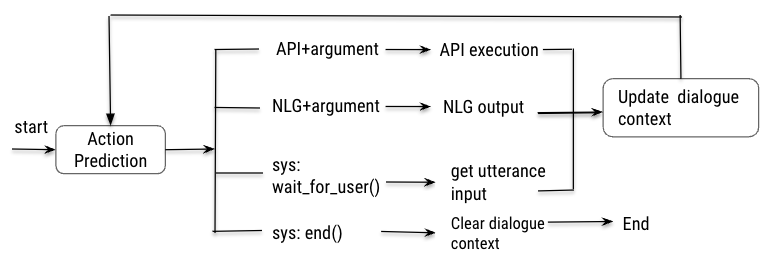}
  \caption{The model prediction flow.}
  \label{fig:model_operation}
  \vspace{-0.5em}
\end{figure}
{\bf Action Prediction} 
The responsibility of prediction modeling is to (a) understand dialogue context so that it can decide which actions should be taken next; (b) fill arguments for selected actions with entities that occurred in dialogue context, with the purpose of guiding the conversation in the loop. In particular, the model predicts the actions in several types, including intent API (e.g., \texttt{SetMyPreference()}) with argument, NLG service with argument (e.g., the corresponding natural language response based on NLG template and its argument, e.g., \texttt{what is a restaurant type}, and Sys system actions (e.g., wait for user input \texttt{wait\_for\_user()} or end the dialog \texttt{end\_dialogue}),  by leveraging the storage of dialogue context and input utterance and context information to the conversational model. Another import module is name entity recognition (NER) model on user utterances so as to restrict the number of entities the model needs to consider in argument filling for action prediction.

{\bf Multi-turn Prediction Loop} Since the model continuously  interacts with
the context storage in the conversation, the context storage is updated at dialog turn level when the model makes a new prediction for the action before handling the control between seeker and provider. For example, the user may request call an intent/API triggered by \texttt{forge everything I have taught you}, and the agent model will perform an API call for searching all users' preferences by \texttt{getAllPreference()}, and then generate one NLG call that informs the user successfully obtaining all user preferences by \texttt{getAllAffinityAction\_success},  before waiting for user input via a sys\_call like \texttt{wait\_for\_user\_input()}. Then after prompting the user to provide the confirmation of the deletion operation via \texttt{ConfirmationEvent()} API call,  the agent will call the intent API of \texttt{ deleteAllAffinityAction(confirmAction)} followed by NLG generation of successful deletion of all preferences via  \texttt{deleteAllAffinityAction\_success}. Fig.~\ref{fig:model_operation} illustrates the model prediction operation loop.

\subsection{Model breakdown}
\label{Section:model_breakdown}

As is illustrated before, the conversation model is not a single monolithic model, but consists of several key models instead. 

{\bf Dialogue Context Encoder} It encodes dialogue context into vector embeddings used by other models. In particular, in dialog contexts, we extract (1) current utterance (denoted as $E_{cu}$), (2) past utterances (denoted as $E_{pu}$), (3) current entities (denoted as $E{ce}$), (4) past entities (denoted as $E_{pe}$),  (5) past actions (denoted as $E_{pa}$) as features, and adopt BERT encoder~\cite{devlin-etal-2019-bert} as embeddings before concatenating them into a unified context embedding $C$, i.e., 
\begin{eqnarray}
        C = [E_{cu}, E_{pu}, E_{ce}, E_{pe}, E_{pa}]. 
\end{eqnarray}

{\bf Action Prediction (AP) Model } Given the embedding of context history,  it predicts $n$-best list of actions (API/NLG/SYS)\footnote{API denotes standard api to achieve user goals, e.g.,  \texttt{get\_all\_preference}, \texttt{set\_all\_preference}, NLG denotes the process of producing natural language output as the response, SYS denotes system related operations, such as \texttt{wait\_for\_user\_input()}, end the conversation, etc} from encoded dialogue context by computing the distribution over possible actions (denoted as $A$) with a multi-layer feed-forward network using ReLu activation and softmax over possible actions, i.e., 
\begin{equation}
    A = \text{Softmax}(\text{ReLu}(W C + b)),
\end{equation}
where $W$ and $b$ are the weight matrix and bias vector learned for action prediction, respectively. An example of action prediction output is shown below:  [(API:get\_all\_preference, 0.75), (API: set\_all\_preference, 0.25]

The responsibility of {\bf Argument Filling (AF) Model} is to fill arguments of predicted actions with entities in dialogues, whereas {\bf NER model}  recognizes the entity in the utterance, which restricts search space for argument filling model.

\subsection{AF model and NER model}
\label{section:AF_NER}

{\bf Argument Filling (AF) Model} It fills arguments of predicted Actions with entities in dialog context.  A complete argument-filled action should be consistent with API/NLG constraints defined in conversation goal descriptions with encoded dialog context. Given each $c_i$ in a representation of the $i$-th token in context encoder, the prediction is a scalar $p_i$ between 0 and 1, indicating the probability of $i$-th token being the argument value. To make a better prediction, the embeddings of \emph{argument\_name}, \emph{argument\_type}, \emph{action\_name} are fed into the model as the dialog context, attention scores are calculated over all hidden states over dialogue context with entities that violates API/NLG constraints masked out. Finally, the highest scoring entity $p_i$ is chosen as the argument of all tokens of $x_i$, i.e., 
\begin{equation}
    p_i = \text{sigmoid}(\text{Attention}(C, x_i)).
\end{equation}


{\bf NER model} It recognizes the entity in the utterance, which restricts search space for argument filling model. 
Given the encoded context history, and the embedding of each (e.g., $i$-th) token in current utterance using BERT,  a bidirectional LSTM-CRF~\cite{DBLP:journals/corr/HuangXY15} is employed to predict the probability vector over all possible entities for inside, outside, beginning tagging by optimizing the objectives: 
\begin{equation}
    Pr(s|x; w) = \frac{exp(w \dot \phi(x,s))}
    {\sum_{s'} exp(w \dot \phi(x,s'))}, 
\end{equation}
where function $\phi()$ maps the input sequence $x$ to state sequence $s$ with model parameter $w$,  and $s'$ ranges over all possible output sequence with the scoring function $w \dot \phi(x,s))$ indicating how well the state sequence fits the given input sequence that is modeled using a bidirectional LSTM with parameter learned by back propagation.


\subsection{NER model optimization via N-gram catalog feature}
\label{appendix:n_gram}
Suppose we have entity label space {\texttt\{sport\_team, city\}}, the \texttt{sport\_team} catalog = \texttt\{San Francisco Giant\}, the \texttt{city} catalog = \texttt\{San Francisco\},  corresponding to each token in the user utterance \texttt{``I follow San Francisco Giant''}, to identify the boundary and type of the entity, we generate the 1-gram catalog feature, i.e., 
$$[ [0,0], [0, 0], [0, 0], [0, 0], [0, 0]],$$ 2-gram catalog feature, i.e., $$[[[0,0], [0, 0], [0, 1], [0, 0], [0, 0]],$$ and 3-gram catalog feature, i.e., $$[[0,0], [0, 0], [1, 0], [0, 0], [0, 0]].$$

\section{Performance Evaluations} 
\label{Section:performance_evaluation}

\subsection{Model Accuracy}  


{\bf Dialogues for training model} To train the model, we generate training dataset 
by sampling user goals (denoted as sequence of APIs, their arguments and values) before generating dialogues from extension of these user goals.  Table~\ref{tab:goldens} (Appendix~\ref{sec:appendix_golden}) lists examples of the goals we would like to support using the corresponding utterances in the dialogue flow. Using dialogue simulation (see Sec~\ref{Section:simulator_data_generation}), 
This totally gives 50,000 dialogues, 546,509 actions and 151,208 turns (Table ~\ref{tab:training}).

{\bf Model performance test on in-sample dialogues} To evaluate the predictive capabilities of the models, 
We generate an in-sample testing dataset using the dialogue simulator, which gives 500 dialogues, 5673 actions and 1507 turns (Table ~\ref{tab:testing_1}).


{\bf Model Robustness testing} To evaluate the robustness of our model, we also collect how users will describe these preferences by starting AMT\footnote{https://www.mturk.com/} task 
to collect the variations of those goal-oriented dialogues 
that would produce utterance level variance (i.e., different utterance for the same meaning) and dialog level variance (i.e., diverse dialogue working flow). Then we manually inspect each dialogue, and remove the redundant and unreasonable ones.  

After collecting these dialogues, to abide by the annotation contract, we manually label these AMT dialogue by properly injecting the dialogues based on dialogue definitions that clearly gives the entity boundaries, action signatures with corresponding parameters in consistent with dialogue generation contract. After data cleaning, finally it gives 500 dialogues and 1480 turns (Table ~\ref{tab:testing}). We call it ``out-of-sample" evaluation dataset since it consists of  utterance-level and dialogue level variations that are \emph{not} covered in training dialogues. 
Compared to in-sample dataset, this dialogue dataset has more variations at both utterance-level and dialogue level, which may not match well with the training dialogues. That's the reason we call it as  ``out-of-sample" evaluation dataset.
The annotated dialogue flows are in the same format that can be processed by conversation understanding models. 
 
  \begin{table}
  \small 
 \caption{Statistics of training dataset}
  \centering
  \begin{tabular}{lccc}
    \toprule
   \#API    & \# Testing dialogues   & \# Action & \# turns  \\
    \midrule
  12        &  50,000  & 546,509 & 151,208   \\
    \bottomrule
  \end{tabular}
  \label{tab:training}
\end{table}
 \begin{table}
 \small
 \caption{Statistics of in-sample evaluation dataset}
  \centering
  \begin{tabular}{lccc}
    \toprule
   \#API    & \# Testing dialogues   & \# Action & \# turns  \\
    \midrule
  12        &  500  & 5673 & 1507   \\
    \bottomrule
  \end{tabular}
  \label{tab:testing_1}
\end{table}
 \begin{table}
 \small
 \caption{Statistics of out-of-sample evaluation dataset}
  \centering
  \begin{tabular}{lccc}
    \toprule
   \#API    & \# Testing dialogues   & \# Action & \# turns  \\
    \midrule
  12        &  500  & 5355 & 1480   \\
    \bottomrule
  \end{tabular}
  \label{tab:testing}
\end{table}



 


{\bf Metrics} We use \emph{turn level accuracy} to measure the performance of the models for the multi-turn dialog system. One turn includes the user's utterance plus the conversation agent’s response, which measures a two-sided conversation in a dialog. 
Each turn is considered as correct when all predictions of models (including NER, AP, AF) in the turn are correct.  For example, for the NER model, we can compare the predictions of all entities against the ground-truth, and define precision, recall, accuracy and F1-score based on the percentage of correctly predicted entities against the ground-truth of entities. The turn level accuracy for AP and AF are similarly defined. We also show the model performance at each action  level (i.e., at API, NLG, SYS levels), and the accuracy is defined based on the percentage of correctly predicted ones against the ground-truth at each-action level. We show model prediction results for each model (such as NER, AP and AF).  Also, we show the end2end model evaluation result (denoted as NER+AP+AF), which indicates the
it is correct if and only if all NER/AP/AF are correct for an action (or a turn). 
 

{\bf Result} Tables~\ref{tab:accuracy_i}, ~\ref{tab:accuracy_o} show both the turn level accuracy and action-level accuracy using both in-sample evaluation dataset and out-of-sample evaluation dataset. The result is expected as the model performs reasonably well on an in-sample dataset, whereas the model did not handle some entities correctly and therefore affects argument filling (AF) model performance as well since these entities might be out of dictionary. 
{\small 
\begin{table}
\small
 \caption{Accuracy (ACC) at per-turn level and per-action level for different models in \emph{in-sample} evaluation dataset}
  \centering
  \begin{tabular}{lcc}
    \toprule
    model     & ACC per-turn   & ACC per-action  \\
    \midrule
    NER        &  99.85\% & 99.85\%   \\
    AP          & 100\% & 100\% \\ 
    AF         & 97.89\% & 98.72\% \\ 
    AP + AF   & 97.89\% & 98.72\%  \\ 
    NER + AP + AF & 97.40\% & 97.65\% \\
    \bottomrule
  \end{tabular}
  \label{tab:accuracy_i}
\end{table}
}
{\small 
\begin{table}
\small
 \caption{Accuracy (ACC) at per-turn level and per-action level for different models in \emph{out-of-sample} evaluation dataset}
  \centering
  \begin{tabular}{lcc}
    \toprule
    model     & ACC per-turn   & ACC per-action  \\
    \midrule
    NER        &  94.48\% & 95.89\%   \\
    AP          & 97.18\% & 97.18\% \\ 
    AF         & 93.92\% & 95.24\% \\ 
    AP + AF   & 92.57\% & 94.92\%  \\ 
    NER + AP + AF & 91.22\% & 91.22\% \\
    \bottomrule
  \end{tabular}
  \label{tab:accuracy_o}
\end{table}
}
\begin{table}
\small
 \caption{Accuracy (ACC) at per-turn level and per-action level for NER models in \emph{out-of-sample} evaluation dataset after adding catalog specific features (denoted as CF)}
  \centering
  \begin{tabular}{lcc}
    \toprule
    model     & ACC per-turn   & ACC per-action  \\
    \midrule
    NER  w/ CF      &  96.79\% & 98.30\%   \\
    NER  w/o CF     &  94.48\% & 95.89\%   \\
    \bottomrule
  \end{tabular}
  \label{tab:accuracy_ner}
\end{table}


\vspace{-2mm}
\subsection{Lessons Learned}
\label{section:lessions_learned}

{\bf Increasing simulated dialogues for addressing cold start issues} Thanks to the dialogue simulator component that generates many training dialogues,  we are able to overcome the difficulty of insufficient training dialogues from cold-start. However, when we did evaluation using AMT data collected from user study, we still observed that some dialogue samples are \emph{not} covered by the simulator. We may leverage the recent advances in generative Q\&A (e.g., GPT-3~\cite{NEURIPS2020_1457c0d6}, OPT-175B model~\cite{https://doi.org/10.48550/arxiv.2205.01068}) to complement current training dialogues from simulators. The training dialogues are augmented by collecting online user dialogues as well after the product is deployed.  All these dialogues are put into \emph{feedback loop} to update the current model. 

{\bf More open and personalized preference suggestions} When we train the model, the entity type is given when setting the user preference. To move forward, some users are interested in receiving more accurate suggestions after setting preferences. We would provide more personalized suggestions based on preferences learned from current users. We may even support open conversations such as \texttt{``What are my favorite \_\_\_"} to see what is the entity list to be added, 
rather than the fixed entity types. 

{\bf Expanding catalogs} When we do offline evaluation for the model, we see NER model performs well on the in-sample evaluation dataset. However, it drops in out-of-sample dataset.  In user study, some users responds wish to be able to add more options to their different categories of preferences.  For example, we may expand sports team options to include college sports and teams, and minor league teams as well. We enabled \emph{n-gram catalog feature} to get catalog specific representations particularly for the entity types that have free-form catalog values which may not be well captured by BERT pre-trained models on Wikipedia data. These catalog features (Appendix~\ref{appendix:n_gram}) are concatenated into the embedding in LSTM-CRF models 
and  Table~\ref{tab:accuracy_ner} shows the performance improvement in out-of-sample dataset for NER tasks. 

\vspace{-2mm}
\subsection{User satisfaction} 
We conducted a user study 
to test the developed feature performance as well. 
Overall, participants expressed satisfaction when teaching their preferences for sports, restaurants and weather providers in a single session, emphasizing how easy and seamless the experience was when setting different types of preferences all at once. These findings are in line with the results observed in model evaluations. 

\begin{figure}[ht!]
\centering
  \includegraphics[scale=0.59]{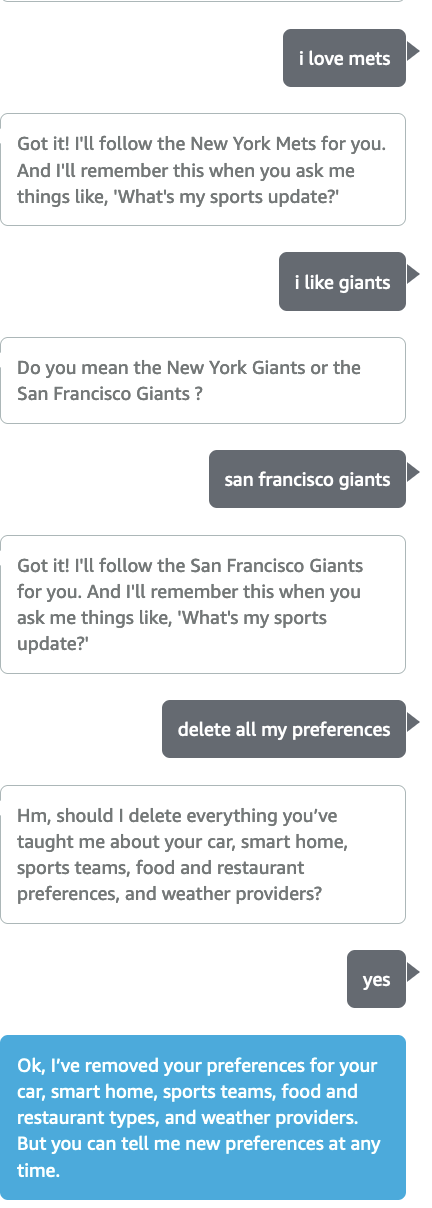}
  \caption{A snapshot for the conversation flow.}
  \label{fig:demon}
\end{figure}

In particular, we notice some reasons for user dis-satisfactions including ASR recognition error, issues with loops/crashes, and incorrect response, which motivates us to further improve the end2end modeling performance.  Fig.~\ref{fig:demon} demonstrates how the user interacts with the conversation agent in the product. 

\begin{table}
\small
 \caption{Top reasons of user dissatisfactions}
  \centering
  \begin{tabular}{lcc}
    \toprule
    Reason    & Ratio(\%)     \\
    \midrule
   Issues with ASR recognition & 25\%  \\
   Issues with loops/errors/crashes       &  20\%   \\
   incorrect/no response          & 25\% \\ 
    Some entities are incorrectly recognized       & 20\% \\ 
    Others  & 10\% \\ 
    \bottomrule
  \end{tabular}
  \label{tab:CAST}
\end{table}



\vspace{80mm}

\vspace{-2mm}
\section{Conclusion}
\label{Section:conclusion}
This paper presents a conversation dialog system capable of teaching users' preference for bridging the gaps of conversation understanding and dialog management. The teachable dialogue system improves the language understanding capabilities and provides more natural conversations to users.  TAI demonstrates good performance in teaching users' preferences in terms of model accuracy and user satisfaction. We are extending the application to various domains by supporting conversation based reasoning and personalized recommendation. 

\bibliographystyle{ACM-Reference-Format}
\bibliography{lmp}




\end{document}